\documentclass[12pt,reqno]{article}
\usepackage{amsfonts}
\usepackage{amsmath}
\usepackage{amsbsy}

\usepackage{amssymb,latexsym}
\numberwithin{equation}{section}

\begin{document}
 \allowdisplaybreaks[1]
\title{Non-Linear Realisation of the Pure $\mathcal{N}=4$, $D=5$ Supergravity}
\author{Nejat T. Y$\i$lmaz\\
Department of Mathematics
and Computer Science,\\
\c{C}ankaya University,\\
\"{O}\u{g}retmenler Cad. No:14,\quad  06530,\\
 Balgat, Ankara, Turkey.\\
          \texttt{ntyilmaz@cankaya.edu.tr}}
\maketitle
\begin{abstract}We perform the non-linear realisation or the coset
formulation of the pure $\mathcal{N}=4$, $D=5$ supergravity. We
derive the Lie superalgebra which parameterizes a coset map whose
induced Cartan-Maurer form produces the bosonic field equations of
the pure $\mathcal{N}=4$, $D=5$ supergravity by canonically
satisfying the Cartan-Maurer equation. We also obtain the
first-order field equations of the theory as a twisted
self-duality condition for the Cartan-Maurer form within the
geometrical framework of the coset construction.

\end{abstract}

\section{Introduction}
The method of dualisation has been used in \cite{julia2} to
formulate the bosonic sectors of the $D=11$ supergravity
\cite{D=11} and the maximal supergravities which are obtained from
the $D=11$ supergravity by torodial compactifications as coset
models. The first-order formulations of these theories are also
embedded in the non-linear coset formulations as twisted
self-duality constraints \cite{julia3}. The main motivation to
study the non-linear realisations of the $D=11$ supergravity and
its Kaluza-Klein descendant theories is to understand the global
symmetries of the $D=11$ supergravity thus the symmetries of the
M-theory.

In general the supergravity theories are the massless sectors or
the low energy effective limits of the relative string theories.
Thus the symmetries of the supergravity theories have been studied
to improve the knowledge of the symmetries and the duality
relations of the string theories. In this respect the global
symmetries of the supergravity theories gain importance since a
restriction of the global symmetry group $G$ of the supergravity
theory to the integers is conjectured to be the U-duality symmetry
of the relative string theory \cite{nej125,nej126}. The non-linear
sigma model or the coset formulation of the supergravities
provides an effective tool to study the global symmetries of these
theories and thus the symmetries of the relative string theories
as mentioned above.

The scalar coset manifolds of the maximal supergravities are based
on split real form global symmetry groups
\cite{julia1,ker1,ker2,nej1,nej2}. In \cite{nej2} the method of
dualisation is extended to the non-split real form scalar cosets.
This enlarged formulation has enabled the coset construction of
the matter coupled supergravities which have non-split scalar
cosets \cite{nej4,nej5,nej6}. A general dualisation and non-linear
realisation is also performed for a generic scalar coset which is
coupled to matter in \cite{nej3}.

In this work we present the non-linear realisation or the coset
construction of the bosonic sector of the pure $\mathcal{N}=4$,
$D=5$ supergravity \cite{d51,d52,d53,d54}. Our formulation will be
in parallel with the ones in \cite{julia2,nej4,nej5,nej6}. We will
propose a coset map and we will derive the Lie superalgebra which
parameterizes the coset representatives so that the Cartan-Maurer
form of the coset map will lead us to the bosonic field equations
of the theory by satisfying the Cartan-Maurer equation. Therefore
the definition of the coset map and the construction of the
algebra which parameterizes this map enables us to formulate the
pure $\mathcal{N}=4$, $D=5$ supergravity as a non-linear coset
sigma model. The first-order formulation of the theory will also
be derived and we will mention its role as a constraint condition
in the coset construction.

In section two we will derive the field equations of the pure
$\mathcal{N}=4$, $D=5$ supergravity. We will also give the locally
integrated first-order field equations. In section three after
introducing the algebra generators and the coset map we will
construct the Lie superalgebra structure of the field generators.
We will also show that the first-order field equations can be
obtained as a twisted self-duality condition within the coset
construction.
\section{The Pure ${\mathcal{N}}=4$, $D=5$ Supergravity}
The field content of the pure $\mathcal{N}=4$, $D=5$ supergravity
\cite{d51,d52,d53,d54} is
\begin{equation}\label{21}
(e_{\mu}^{r},\:\psi_{\mu}^{i},\:\chi^{j},\: v_{\mu},\:
V_{\mu}^{k},\:\phi),
\end{equation}
where $e_{\mu}^{r}$ is the f\"{u}nfbein, $v_{\mu}$ is a one-form
field in the $\bf{1}$ representation of $USp(4)$, $V_{\mu}^{k}$
are five one-form fields for $k=1,...,5$ in the $\bf{5}$
representation of $USp(4)$, $\phi$ is a scalar field which is
singlet under the action of $USp(4)$, also $\psi_{\mu}^{i}$ for
$i=1,...,4$ are four gravitini in the $\bf{4}$ representation of
$USp(4)$ and for $j=1,...4$ $\chi^{j}$ are four spin-$1/2$ fields
again in the $\bf{4}$ representation of $USp(4)$. The local
symmetries of the pure ${\mathcal{N}}=4$, $D=5$ supergravity are
the general coordinate transformations, ${\mathcal{N}}=4$
supersymmetry and the $U(1)^{6}$ gauge invariance whose abelian
gauge fields are the six one-forms. The global symmetry of the
theory or the U-duality group is $USp(4)\times SO(1,1)$ which is
an electric subgroup of the $D=4$ $Sp(14,\bf{R})$ duality group.
The R-symmetry group $USp(4)$ is the automorphism group of the
${\mathcal{N}}=4$, $D=5$ supersymmetry algebra and $USp(4)$ is
isomorphic to $Spin(5)$ which is the covering group of $SO(5)$.
For the following formulation in accordance with \cite{d54} the
signature of the space-time metric will be chosen as
\begin{equation}\label{215}
\eta_{AB}=\text{diag}(-,+,+,+,+).
\end{equation}
Apart from the gravity sector the bosonic lagrangian of the pure
$\mathcal{N}=4$, $D=5$ supergravity can be given as
\cite{d51,d52,d53,d54}
\begin{equation}\label{22}
\begin{aligned}
\mathcal{L}&=-\frac{1}{2}\, \ast d\phi\wedge d\phi-\frac{1}{2}\,
e^{\frac{2}{\sqrt{6}}\phi}\,\ast dV^{i}
\wedge dV_{i}\\
\\
&\quad -\frac{1}{2}\, e^{-\frac{4}{\sqrt{6}}\phi}\,\ast dv\wedge
dv-\frac{1}{2}\, dV^{i} \wedge dV_{i}\wedge v,
\end{aligned}
\end{equation}
where $i=1,...,5$ and we should remark that we raise or lower the
indices by using the Euclidean metric. If we vary the lagrangian
\eqref{22} with respect to the fields $\phi$, $v$, $V^{i}$ we can
find the second-order field equations respectively as
\begin{subequations}\label{23}
\begin{gather}
d(\ast d\phi)=\frac{1}{\sqrt{6}}\,
e^{\frac{2}{\sqrt{6}}\phi}\,\ast dV^{i}\wedge
dV_{i}-\frac{2}{\sqrt{6}}\, e^{-\frac{4}{\sqrt{6}}\phi}\,\ast
dv\wedge dv,\notag\\
\notag\\
d(e^{-\frac{4}{\sqrt{6}}\phi}\,\ast dv)=
-\frac{1}{2}\, dV^{i}\wedge dV_{i},\notag\\
\notag\\
d(e^{\frac{2}{\sqrt{6}}\phi}\,\ast dV^{j}+dV^{j}\wedge
v)=0.\tag{\ref{23}}\end{gather}
\end{subequations}
We will locally integrate the bosonic field equations \eqref{23}
by introducing dual fields and by using the fact that locally a
closed form is an exact one. Integration will give us the
first-order field equations of the theory. Here by integration we
mean to cancel an exterior derivative on both sides of the
equations. Thus if we introduce the three-form dual field
$\widetilde{\phi}$ for the original field $\phi$, the five
two-form fields $\widetilde{V}^{i}$ for $V^{i}$ and the two-form
field $\widetilde{v}$ for $v$ we can express the first-order field
equations obtained from the second-order field equations
\eqref{23} as explained above as
\begin{subequations}\label{24}
\begin{gather}
\ast d\phi=-\frac{1}{\sqrt{6}}\, V^{i}\wedge
d\widetilde{V}_{i}+\frac{2}{\sqrt{6}}\, v\wedge d\widetilde{v}- d\widetilde{\phi},\notag\\
\notag\\
e^{\frac{2}{\sqrt{6}}\phi}\,\ast dV^{i}
+V^{i}\wedge dv=-d\widetilde{V}^{i},\notag\\
\notag\\
e^{-\frac{4}{\sqrt{6}}\phi}\,\ast dv=-d\widetilde{v}-\frac{1}{2}\,
V^{i}\wedge dV_{i}.\tag{\ref{24}}
\end{gather}
\end{subequations}
When one applies the exterior derivative on both sides of the
equations above one obtains the second-order field equations given
in \eqref{23} which do not contain the dual fields
$\widetilde{\phi}$, $\widetilde{v}$, $\widetilde{V}^{i}$. In the
next section when we give the non-linear coset formulation of the
bosonic sector of the pure $\mathcal{N}=4$, $D=5$ supergravity we
will show that the first-order field equations in \eqref{24} can
be obtained from a twisted self-duality condition \cite{julia2}
which the Cartan-Maurer form of the coset map satisfies.
\section{The Coset Formulation}

In this section we will present the coset construction or the
non-linear realisation of the bosonic sector of the pure
$\mathcal{N}=4$, $D=5$ supergravity. We will introduce the coset
map whose Cartan form realizes the second-order bosonic field
equations \eqref{23} by satisfying the Cartan-Maurer equation. Our
main objective will be to derive the Lie superalgebra which
parameterizes the coset elements and which leads to the non-linear
coset formulation of the theory. We will follow the method of
dualisation or the doubled formalism of \cite{julia2} to define
the coset map and to obtain the necessary commutation and the
anti-commutation relations of the generators of the Lie
superalgebra which parameterizes the coset representatives.

Likewise in \cite{julia2,nej4,nej5,nej6} as a first task we
introduce a dual field for each original bosonic field given in
\eqref{21} namely for the fields $(v$, $V^{k}$, $\phi)$. The
corresponding dual fields are
\begin{equation}\label{31}
(\,\widetilde{v},\,\widetilde{V}^{k},\,\widetilde{\phi}\,),
\end{equation}
where $\widetilde{\phi}$ is a three-form, for $k=1,...,5\,$
$\widetilde{V}^{k}$ are two-forms also $\widetilde{v}$ is a
two-form field. It will be clear later when we obtain the
first-order equations from the explicit calculation of the
Cartan-Maurer form of the constructed coset map that these fields
coincide with the fields which we have already introduced in the
integrated field equations in \eqref{24}. As a matter of fact
these fields are the necessary langrange multipliers to construct
the Bianchi lagrangians from the Bianchi identities of the
corresponding original fields \cite{pope}. Next we assign a
generator for each bosonic field and its dual field. The original
generators are
\begin{equation}\label{32}
(\,Y,\, Z_{k},\, K\,),
\end{equation}
which in the coset map will be coupled to the fields $(v$,
$V^{k}$, $\phi)$ respectively. Besides, the dual generators which
will be coupled to the dual fields $\widetilde{v},$
$\widetilde{V}^{i}$, $\widetilde{\phi}$ in the coset map will be
defined as
\begin{equation}\label{33}
(\,\widetilde{Y},\,\widetilde{Z}_{i},\,\widetilde{K}\,),
\end{equation}
respectively. The Lie superalgebra generated by the generators in
\eqref{32} and \eqref{33} will have the $\mathbb{Z}_{2}$ grading.
The generators will be defined to be odd if the corresponding
coupling field is an odd degree differential form and otherwise
even \cite{julia2}. Therefore the generators
$\{Z_{i},\,Y,\,\widetilde{K}\}$ are the odd generators and
$\{K,\,\widetilde{Z}_{i},\,\widetilde{Y}\}$ are the even ones. The
coset map which we have mentioned above will be parameterized by a
differential graded algebra \cite{julia2}. This algebra contains
the local module of the differential forms and the Lie
superalgebra of the original and the dual generators we have
introduced above. The differential graded algebra structure has
the property that the odd (even) generators behave like odd (even)
degree differential forms when they commute with the exterior
product. The algebra products of the odd generators obey the
anti-commutation relations whereas the algebra products of the
even ones and the mixed ones obey the commutation relations within
the Lie superalgebra structure.

Equipped with the necessary algebraic tools, to start the
construction of the coset realisation of the pure $\mathcal{N}=4$,
$D=5$ supergravity we first introduce the coset map
\begin{equation}\label{34}
\nu^{\prime}=e^{\phi K}e^{V^{i}Z_{i}}e^{vY}
e^{\widetilde{v}\widetilde{Y}}e^{\widetilde{V}^{i}\widetilde{Z}_{i}}e^{\widetilde{\phi}
\widetilde{K}}.
\end{equation}
This parametrization may be considered as a map from the
five-dimensional space-time into a symmetry group whose structure
will not be the interest of this work. Like in
\cite{julia2,nej4,nej5,nej6} we will solely focus on the
derivation of the Lie superalgebra which generates the coset
parametrization \eqref{34}. However certainly this algebra would
reflect the properties of the target group at least locally
\cite{hel}. In a chosen matrix representation of the Lie
superalgebra the pullback of the Cartan-Maurer form
$\mathcal{G}^{\prime}$ which is defined on the above mentioned
target group $G$ through the map \eqref{34} can be given as
\begin{equation}\label{35}
\mathcal{G}^{\prime}=d\nu^{\prime}\nu^{\prime-1}.
\end{equation}
As a result of its definition the Cartan form \eqref{35} satisfies
the Cartan-Maurer equation
\begin{equation}\label{36}
d\mathcal{G}^{\prime}-\mathcal{G}^{\prime}\wedge\mathcal{G}^{\prime}=0.
\end{equation}
Our construction will be based on the requirement that the
Cartan-Maurer equation should produce the second-order field
equations \eqref{23} when calculated explicitly. At first glance
one may observe that to calculate the Cartan-Maurer form thus the
Cartan-Maurer equation explicitly one needs the algebra
commutators and the anti-commutators of the field generators
\eqref{32} and \eqref{33}. Therefore one has to discover the Lie
superalgebra structure which would give the correct second-order
bosonic field equations \eqref{23} via the Cartan-Maurer equation.
When one reaches the algebra structure one completes the coset
construction and succeeds in the non-linear realisation of the
bosonic sector of the pure $\mathcal{N}=4$, $D=5$ supergravity.
The construction of the algebra of the field generators will
enable us to express the bosonic field equations as ingredients of
a coset structure. The direct but the cumbersome way of solving
the algebra structure is to calculate the Cartan-Maurer equation
\eqref{36} in terms of the unknown structure constants of the
algebra and then to read the structure constants by comparing the
result with the second-order equations given in \eqref{23}. To
advance in this direction one needs to make use of the matrix
identities
\begin{equation}\label{37}
\begin{aligned}
de^{X}e^{-X}&=dX+\frac{1}{2!}[X,dX]+\frac{1}{3!}[X,[X,
dX]]+....,\\
\\
e^{X}Ye^{-X}&=Y+[X,Y]+\frac{1}{2!}[X,[X,Y]]+....,
\end{aligned}
\end{equation}
repeatedly starting from the definition of the Cartan-Maurer form
\eqref{35} and then one should insert the result in the
Cartan-Maurer equation \eqref{36} \cite{julia2,julia1}. We should
remark one point again, like in \cite{nej4,nej5,nej6} we assume
that we work in a matrix representation of the algebra of the
original and the dual generators. We will not lay out the long
steps of the calculation discussed above. Thus we will only
present the results we have obtained. If one performs the above
mentioned calculation one finds that the proper commutation and
the anti-commutation relations of the original and the dual
generators in \eqref{32} and \eqref{33} which lead to the
second-order field equations of \eqref{23} in \eqref{36} are
\begin{subequations}\label{38}
\begin{gather}
[K,Z_{i}]=\frac{1}{\sqrt{6}}\: Z_{i}\quad,\quad[K,Y]=-\frac{2}{\sqrt{6}}\:Y\quad,\quad[K,\widetilde{Y}]=\frac{2}{\sqrt{6}}\:\widetilde{Y},\notag\\
\notag\\
[K,\widetilde{Z}_{i}]=-\frac{1}{\sqrt{6}}\:\widetilde{Z}_{i}\quad,\quad\{Z_{i},Z_{j}\}=\delta_{ij}\:\widetilde{Y}\quad,\quad
\{Z_{i},Y\}=\widetilde{Z}_{i},\notag\\
\notag\\
[\widetilde{Z}_{i},Z_{j}]=\frac{1}{\sqrt{6}}\:\delta_{ij}\:\widetilde{K}
\quad,\quad[\widetilde{Y},Y]=-\frac{2}{\sqrt{6}}\:\widetilde{K}.\tag{\ref{38}}
\end{gather}
\end{subequations}
The commutation and the anti-commutation relations which are not
listed in \eqref{38} vanish. If $O$ represents the set of the
original generators and $\widetilde{D}$ the set of the dual
generators we observe that the algebra constructed in \eqref{38}
obeys the general structure
\begin{subequations}\label{39}
\begin{gather}
[O,\widetilde{D}\}\subset\widetilde{D}\quad,\quad
[\widetilde{D},\widetilde{D}\}=0.\tag{\ref{39}}
\end{gather}
\end{subequations}
However unlike the algebras constructed in \cite{nej4,nej5,nej6}
we have
\begin{equation}\label{310}
[O,O\}\subset O\cup\widetilde{D}.
\end{equation}
Now since we have derived the algebra structure of the field
generators we can calculate the Cartan-Maurer form
$\mathcal{G}^{\prime}=d\nu^{\prime}\nu^{\prime-1}$ starting from
the coset map \eqref{34} exactly. The calculation again needs the
use of the identities in \eqref{37} together with the algebra
structure derived in \eqref{38}. The calculation yields

\begin{equation}\label{311}
\begin{aligned}
\mathcal{G}^{\prime}&=d\nu^{\prime}\nu^{\prime-1}\\
\\
&=d\phi\,
K\: +\:e^{\frac{1}{\sqrt{6}}\phi}\,dV^{i}\,Z_{i}\: +\:e^{-\frac{2}{\sqrt{6}}\phi}\,dv\,Y\\
\\
&\quad
+(\:\frac{1}{2}\,e^{\frac{2}{\sqrt{6}}\phi}\,\delta_{ij}\,V^{i}\,\wedge\,
dV^{j}\: +\:e^{\frac{2}{\sqrt{6}}\phi}\,d\widetilde{v}\: )\,\widetilde{Y}\\
\\
&\quad +(\: e^{-\frac{1}{\sqrt{6}}\phi}\,V^{i}\,\wedge\, dv\:
+\:e^{-\frac{1}{\sqrt{6}}\phi}\,d\widetilde{V}^{i}\:)\,\widetilde{Z}_{i}\\
\\
&\quad +(\:d\widetilde{\phi}\:
+\:\frac{1}{\sqrt{6}}\,V^{i}\,\wedge\, d\widetilde{V}_{i}\:
-\:\frac{2}{\sqrt{6}}\,v\,\wedge\, d\widetilde{v}\:
)\,\widetilde{K}.
\end{aligned}
\end{equation}
The non-linear realisation of the supergravity theories as a
result of the dualisation of the fields is intimately related to
the langrange multiplier methods which give rise to the
first-order formulations of these theories \cite{pope}. For this
reason our coset formulation also yields the locally integrated
first-order field equations \eqref{24}. To obtain the first-order
field equations from the Cartan-Maurer form \eqref{311} we have to
introduce a pseudo-involution $\mathcal{S}$ of the Lie
superalgebra constructed in \eqref{38}. By following the general
scheme introduced in \cite{julia2} also discussed in
\cite{nej1,nej2} we can define the action of $\mathcal{S}$ on the
field generators as
\begin{subequations}\label{312}
\begin{gather}
\mathcal{S}Y=\widetilde{Y}\quad,\quad\mathcal{S}K=\widetilde{K}\quad,\quad
\mathcal{S}Z_{i}=\widetilde{Z}_{i},\notag\\
\notag\\
\mathcal{S}\widetilde{Y}=-Y\quad,\quad\mathcal{S}\widetilde{K}=-K\quad,\quad
\mathcal{S}\widetilde{Z}_{i}=-Z_{i}.\tag{\ref{312}}
\end{gather}
\end{subequations}
Now if we require the Cartan-Maurer form in \eqref{311} to obey
the twisted self-duality constraint
\begin{equation}\label{313}
\ast\mathcal{G}^{\prime}=\mathcal{SG}^{\prime},
\end{equation}
we observe that by equating the coefficients of the linearly
independent algebra generators in \eqref{313} when \eqref{311} is
inserted in it we get exactly the first-order field equations
derived in \eqref{24}. Since the equations in \eqref{24} have been
obtained from the second-order field equations \eqref{23}
algebraically owing to abolishing an exterior derivative locally
we conclude that the twisted self-duality constraint \eqref{313}
which is reserved on the Cartan-Maurer form \eqref{311} is a
justified condition. Thus our formulation have not only
re-established the bosonic sector of the pure $\mathcal{N}=4$,
$D=5$ supergravity as a coset model but it also has produced the
first-order field equations as a twisted self-duality condition
within the coset construction.
\section{Conclusion}
We have applied the formalism of dualisation to construct the
non-linear coset formulation of the bosonic sector of the pure
$\mathcal{N}=4$, $D=5$ supergravity \cite{d51,d52,d53,d54}. We
have introduced a coset map and constructed the Lie superalgebra
which parameterizes this map so that the Cartan-Maurer form
induced by the coset map generates the second-order bosonic field
equations of the theory in the canonical Cartan-Maurer equation.
In this way we have established the necessary algebraic background
to interpret the bosonic sector of the pure $\mathcal{N}=4$, $D=5$
supergravity as a non-linear coset sigma model. The bosonic field
equations are regained as elements of the geometrical framework of
the coset construction. As a consequence of the dualisation method
which originates from \cite{julia2} the first-order formulation of
the theory is also obtained in the coset formulation. The
first-order field equations appear as a twisted self-duality
constraint \cite{julia2,julia3} which the Cartan-Maurer form
satisfies. Therefore we get the first-order field equations as a
byproduct of our formulation. This is not a surprise as the method
of dualisation is another manifestation of the lagrange multiplier
formalism.

Although we have derived the Lie superalgebra which forms the
gauge to parameterize the coset representatives in our non-linear
construction we have not inquired the group theoretical structure
of the coset. The coset map can be considered to be into a group
$G$ which is presumably \cite{julia2,julia1} the global symmetry
group of the doubled lagrangian which is obtained as a result of
the dualisation of all the fields. Thus the study of the coset
structure may contribute to the understanding of the global
symmetries of the theory. The local symmetries can also be
examined in the same way. The Lie superalgebra we have constructed
can be considered as a key and a good starting point in the
identification of the coset structure \cite{hel}.

One may couple matter multiplets to the pure $\mathcal{N}=4$,
$D=5$ supergravity studied in this work \cite{d51,d52,d53,d54}.
The resulting Maxwell-Einstein supergravity can be related to the
five-dimensional Kaluza-Klein descendant theory of the
ten-dimensional low energy effective heterotic string by the
redefinition, truncation and dualisation of appropriate fields
\cite{d52,heterotic}. Therefore the coset formulation of this work
which brings an insight in the symmetry structure may also help to
understand the symmetries of the heterotic string.

The gravity and the fermionic sectors can also be included in the
non-linear realisation analysis. Especially the coset formulation
can be extended to include the gravity sector in accordance with
\cite{west1,west2,west3} so that the Kac-Moody symmetries of the
pure $\mathcal{N}=4$, $D=5$ supergravity can be studied. One may
also inspect the comparison of the Lie superalgebra derived here
with the ones constructed in \cite{nej4,nej5,nej6} to draw a
general scheme of symmetries.


\begin{thebibliography}{99}
\bibitem{julia2}
 E. Cremmer, B. Julia, H. L\"{u} and C. N. Pope, ``\textit{Dualisation of dualities II : Twisted self-duality of doubled fields and
 superdualities}",
Nucl. Phys. \textbf{B535} (1998) 242, hep-th/9806106.
\bibitem{D=11}
E. Cremmer, B. Julia and J. Scherk, ``\textit{Supergravity theory
in eleven-dimensions}", Phys. Lett. \textbf{B76} (1978) 409.
\bibitem{julia3}
B. Julia, ``\textit{Superdualities: below and beyond U-duality}",
LPTENS \textbf{00/02} (2000), hep-th/0002035.
\bibitem{nej125}
C. M. Hull and P. K. Townsend, ``\textit{Unity of superstring
dualities}", Nucl. Phys. \textbf{B438} (1995) 109, hep-th/9410167.
\bibitem{nej126}
E. Witten, ``\textit{String theory dynamics in various
dimensions}", Nucl. Phys. \textbf{B443} (1995) 85, hep-th/9503124.
\bibitem{julia1} E. Cremmer, B. Julia,
H. L\"{u} and C. N. Pope, ``\textit{Dualisation of dualities.
I.}", Nucl. Phys. \textbf{B523} (1998) 73, hep-th/9710119.
\bibitem{ker1}
A. Keurentjes, ``\textit{The group theory of oxidation}", Nucl.
Phys. \textbf{B658} (2003) 303, hep-th/0210178.
\bibitem{ker2}
A. Keurentjes, ``\textit{The group theory of oxidation II : Cosets
of non-split groups}", Nucl. Phys. \textbf{B658} (2003) 348,
hep-th/0212024.
\bibitem{nej1}
N. T. Y$\i$lmaz, ``\textit{Dualisation of the general scalar coset
in supergravity theories}", Nucl. Phys. \textbf{B664} (2003) 357,
hep-th/0301236.
\bibitem{nej2}
N. T. Y$\i$lmaz, ``\textit{The non-split scalar coset in
supergravity theories}", Nucl. Phys. \textbf{B675} (2003) 122,
hep-th/0407006.
\bibitem{nej4}
T. Dereli and N. T. Y$\i$lmaz, ``\textit{Dualisation of the
Salam-Sezgin d=8 supergravity }", Nucl. Phys. \textbf{B691} (2004)
223, hep-th/0407004.
\bibitem{nej5}
N. T. Y$\i$lmaz, ``\textit{Dualisation of the $d=7$ heterotic
string}", JHEP \textbf{0409} (2004) 003, hep-th/0507008.
\bibitem{nej6}
N. T. Y$\i$lmaz, ``\textit{Dualisation of the $d=9$ matter coupled
supergravity}", JHEP \textbf{0506} (2005) 031, hep-th/0507011.
\bibitem{nej3}
T. Dereli and N. T. Y$\i$lmaz, ``\textit{Dualisation of the
symmetric space sigma model with couplings}", Nucl. Phys.
\textbf{B705} (2005) 60, hep-th/0507007.
\bibitem{d51}
E. Cremmer, ``\textit{Supergravities in 5 dimensions}", LPTENS
\textbf{80/17}, (Invited paper at the Nuffield Gravity Workshop,
Cambridge, Eng., Jun 22 - Jul 12, 1980).
\bibitem{d52}
M. Awada and P. K. Townsend, ``\textit{N=4 Maxwell-Einstein
supergravity in five-dimensions and its SU(2) gauging}", Nucl.
Phys. \textbf{B255} (1985) 617.
\bibitem{d53}
G. Dall'Agata, C. Herrmann and M. Zagermann, ``\textit{General
matter coupled N = 4 gauged supergravity in five dimensions}",
Nucl. Phys. \textbf{B612} (2001) 123, hep-th/0103106.
\bibitem{d54}
G. Villadoro and F. Zwirner, ``\textit{The minimal N = 4 no-scale
model from generalized dimensional reduction}", JHEP \textbf{0407}
(2004) 055, hep-th/0406185.
\bibitem{pope}
C. N. Pope, ``\textbf{Lecture Notes on Kaluza-Klein Theory}",
(unpublished).
\bibitem{hel}
S. Helgason, ``\textbf{Differential Geometry, Lie Groups and
Symmetric Spaces}", (Graduate Studies in Mathematics 34, American
Mathematical Society Providence R. I. 2001).
\bibitem{heterotic}
H. L\"{u}, C. N. Pope and K. S. Stelle,
``\textit{M-theory/heterotic duality: A Kaluza-Klein
perspective}", Nucl. Phys. \textbf{B548} (1999) 87,
hep-th/9810159.
 \bibitem{west1}
P. West, ``\textit{Hidden superconformal symmetry in M theory}",
JHEP \textbf{0008} (2000) 007, hep-th/0005270.
\bibitem{west2}
P. West, ``\textit{E(11) and M theory}", Class. Quant. Grav.
\textbf{18} (2001) 4443, hep-th/0104081.
\bibitem{west3}
I. Schnakenburg and P. West, ``\textit{Kac-Moody symmetries of IIB
supergravity}", Phys. Lett. \textbf{B517} (2001) 421,
hep-th/0107181.
\end{thebibliography}
\end{document}